\documentclass[10pt, two column, twoside]{IEEEtran}
\usepackage{cite}

\usepackage{amssymb}
\usepackage{amsmath}
\usepackage[lined,boxed,commentsnumbered, ruled]{algorithm2e}
\usepackage{mathrsfs}
\usepackage{algorithmic}
\usepackage{bm}
\usepackage{tikz}
\usetikzlibrary{arrows}
\usepackage{subfig}
\usepackage{graphicx,booktabs,multirow}

\definecolor{color1}{HTML}{D0B22B}
\definecolor{dred}{RGB}{128,0,0}
\definecolor{colorhkust}{RGB}{20,43,140}
\definecolor{colorshanghaitech}{RGB}{162,0,5}
\definecolor{colortsinghua}{RGB}{116,52,129}
\definecolor{colordark}{RGB}{184,134,11}
\usepackage{amsthm}
\theoremstyle{definition}

\newtheorem{proposition}{Proposition}
\newtheorem{definition}{Definition}

\newcommand{\bs}[1]{{\bm{#1}}}

\usepackage{array}
\newcolumntype{C}[1]{>{\centering\arraybackslash$}m{#1}<{$}}
\newlength{\mycolwd}                                         
\begin{document}
\title{Data Shuffling in Wireless Distributed Computing via Low-Rank Optimization}
\author{Kai~Yang,~\IEEEmembership{Student Member,~IEEE,}
        Yuanming~Shi,~\IEEEmembership{Member,~IEEE,}
        and~Zhi~Ding,~\IEEEmembership{Fellow,~IEEE}

\thanks{Part of this paper was presented at the IEEE International Conference on Acoustics, Speech, and Signal Processing, Calgary, Alberta, Canada, Apr. 2018 \cite{yang2016wdc_icassp}.}
\thanks{K. Yang is with the School of Information Science and Technology, ShanghaiTech University, Shanghai, China, and also with the University of Chinese Academy of Sciences, Beijing, China (e-mail: yangkai@shanghaitech.edu.cn).}
\thanks{Y. Shi is with the School of Information Science and Technology, ShanghaiTech University, Shanghai, China (e-mail: shiym@shanghaitech.edu.cn).} \thanks{Z. Ding is with the Department of Electrical and Computer Engineering,
University of California at Davis, Davis, CA 95616 USA (e-mail:
zding@ucdavis.edu).}
}
\maketitle

\begin{abstract}
Intelligent mobile platforms such as smart vehicles and drones have recently become the focus of attention for onboard deployment of machine learning mechanisms to enable low latency decisions with low risk of privacy breach. However, most such machine learning algorithms are both computation-and-memory intensive, which makes it highly difficult to implement the requisite computations on a single device of limited computation, memory, and energy resources. Wireless distributed computing presents new opportunities by pooling the computation and storage resources among devices. For low-latency applications, the key bottleneck lies in the exchange of intermediate results among mobile devices for \emph{data shuffling}. To improve communication efficiency, we propose a co-channel communication model and design transceivers by exploiting the locally computed intermediate values as side information. A low-rank optimization model is proposed to maximize the achieved degrees-of-freedom (DoF) by establishing the interference alignment condition for data shuffling. Unfortunately, existing approaches to approximate the rank function fail to yield satisfactory performance due to the poor structure in the formulated low-rank optimization problem. In this paper, we develop an efficient DC algorithm to solve the presented low-rank optimization problem by proposing a novel DC representation for the rank function. Numerical experiments demonstrate that the proposed DC approach can significantly improve the communication efficiency whereas the achievable DoF almost remains unchanged when the number of mobile devices grows.
\end{abstract}



%

\section{Introduction}
The mass use of smart mobile devices and Internet-of-Things (IoT) devices promotes the prosperity of mobile applications, and also poses great opportunities for mobile edge intelligence thanks to large amounts of collected input data from end devices. Machine learning has become a key enabling technology for big data analytics and diverse artificial intelligence applications, including computer vision and natural language processing. Increasingly, more and more machine learning applications are executing real-time and private tasks on mobile devices, such as augmented reality, smart vehicles, and drones. However, the ultra-low latency requirement \cite{sze2017efficient} for executing intensive computation tasks of mobile edge intelligence applications imposes an unrealistic burden on the computational capability of resource-constrained mobile devices \cite{han2015deep} and ranks as one of the key challenges. Given limited resources of computation, storage and energy at mobile devices,  a single device often cannot execute the various computation tasks required in learning and artificial intelligence. Wireless distributed computing \cite{Li2017ascalable} promises to support computation intensive intelligent tasks execution on end devices by pooling the computation and storage resources of multiple devices.

Storage size is often one of the key limiting factors in a single device when deploying deep learning model \cite{han2015deep,cheng2018model}. In wireless distributed computing systems for large-scale intelligent tasks, the dataset (e.g., a feature library of objects) is normally too large to be stored in a single mobile device. In popular distributed computing framework such as MapReduce \cite{dean2008mapreduce}, the dataset shall be split and stored across devices in advance, during the \emph{dataset placement phase}. For focal scenarios where each mobile user collects its own input data (e.g., feature vector of an image) and requests the output of its computation task (e.g., inference result of the image), each mobile device shall perform local computation according to locally stored dataset, which is called the \emph{map phase}. Next, in the \emph{shuffle phase}, the computed intermediate values in map phase are exchanged among devices, the output of each mobile device can be constructed with additional local computations (i.e., \emph{reduce phase}). To enable real-time and low-latency applications, inter-device communications for data shuffling in distributed computing system become the main bottleneck. 

To reduce the communication load for data shuffling in distributed computing system, many efforts have focused on designing coded shuffling strategies. The authors of \cite{li2017fundamental} exploited the coded multicast opportunities by proposing a coded scheme called ``Coded MapReduce'' to reduce the communication load for data shuffling in wireline distributed computing framework. In \cite{Li2017ascalable}, a scalable framework for wireless distributed computing is designed, where mobile devices are connected to a common access point (AP) such that the data shuffling is accomplished through orthogonal uplink transmission and via broadcasting at the rate of weakest user on downlink transmission. In this communication model, a coding scheme is proposed to reduce the \emph{communication load} (i.e., \emph{the number of information bits}) for data shuffling. However, in wireless networks with limited spectral resources and interference, it is also critical to improve the \emph{communication efficiency} (i.e., \emph{achieved data rates}) for data shuffling. In this paper, we propose a systematic linear coding approach to improve the communication efficiency in the shuffle phase. To improve spectral efficiency, we assume co-channel transmission in both uplink and downlink. By exploiting the locally computed intermediate values in the map phase as side information, we propose to utilize the interference alignment \cite{cadambe2008interference} (IA) technique for transceiver design in data shuffling.

By establishing the IA condition for data shuffling, we further develop a low-rank model to maximize the achievable degrees-of-freedom (DoF), i.e., the first-order characterization for the achievable data rate. Low-rank approaches have attracted enormous attention in machine learning, high-dimensional statistics, and recommendation system \cite{davenport2016overview}. Unfortunately, the non-convexity of rank function makes the resulting low-rank optimization problem highly intractable. A growing volumn of research focuses on finding tractable approximations for the rank function and on developing efficient algorithms. In particular, nuclear norm relaxation approach is well-known as the convex surrogate of rank function \cite{davenport2016overview}. However, with poorly structured affine constraints in the proposed low-rank optimization model, convex relaxation approach fails to yield satisfactory performance. To further improve the performance of nuclear norm relaxation and enhance low-rankness, the iterative reweighted least square algorithm IRLS-$p$ \cite{mohan2012iterative} ($0\leq p \leq 1$) is proposed by alternating between minimizing weighted Frobenius norm and updating weights. However, such approach still yields unsatisfactory performance under poorly structured affine constraint, which motivates tight and computationally feasible approximations for the rank function. Recently, a DC (difference-of-convex-functions) \cite{tao1997convex,le2018dc} representation of the rank function has been proposed in \cite{gotoh2017dc} with demonstrated effectiveness. Unfortunately, during each iteration of the DC approach, a nuclear norm minimization problem needs to be solved in terms of a semidefinite program and does not scale well to large problem sizes for the data shuffling problem in wireless distributed computing. Motivated by the various issues in the state-of-the art, we shall propose a novel DC approach which is computation efficient and applicable for wireless distributed computing scenario.

\subsection{Contributions}
In this paper, we propose a co-channel communication model for the data shuffling problem in wireless distributed computing system to improve the communication efficiency. Under this model, we adopt linear coding scheme and establish the interference alignment condition for data shuffling. Furthermore, we propose a low-rank optimization model for transceiver design to support efficient algorithms design. To optimize the transceivers with the proposed low-rank model, we propose a novel DC representation for rank function. Specifically, we observe that if the rank of a matrix is $k$, its Ky Fan $2$-$k$ norm should be equal to its Frobenius norm. By alternatively increasing rank and minimizing the difference between the square of Frobenius norm and the square of Ky Fan $2$-$k$ norm, we develop a novel DC approach for the presented low-rank optimization problem. The Frobenius norm allows us to further derive the closed-form solution for each iteration. During each iteration only a subspace projection needs to be computed.

The major contributions of this work are summarized as follows:
\begin{enumerate}
  \item We propose a co-channel communication model for the data shuffling problem in wireless distributed computing. We adopt linear coding scheme in this work, and establish the interference alignment condition for transceiver design. A low-rank model is then developed to maximize the achievable DoF satisfying interference alignment conditions.
  \item To improve communication efficiency, we develop a novel computationally efficient DC algorithm for the low-rank optimization problem. This is achieved by proposing a novel DC representation for rank function. The proposed DC algorithm converges to critical points from arbitrary initial points. 
  \item Numerical experiments demonstrate that with the proposed communication model and DC algorithm, data shuffling in wireless distributed computing can be accomplished with high communication efficiency. The proposed DC algorithm significantly outperforms the nuclear norm relaxation approach and the IRLS algorithm. Furthermore, the communication efficiency is scalable to the number of mobile devices.
\end{enumerate}
This work proposes a systematic framework for efficient data shuffling in wireless distributed computing.
\subsection{Organization and Notation}
The rest of this work is organized as follows. Section II describes the system model of wireless distributed computing, including the computation model and the proposed communication model. Section III provides the interference alignment conditions for data shuffling as well as the formulated low-rank model. Section IV introduces our proposed DC approach. We conduct numerical experiments and illustrate the performance of the proposed algorithm and other state-of-art algorithms in Section V before concluding this work in Section VI.

We use $[N]$ to denote the set $\{1,\cdots,N\}$ for some positive integer $N$. $\otimes$ is the Kronecker product operator. The cardinality of a set $\mathcal{F}$ is denoted by $|\mathcal{F}|$. $\textrm{det}(\cdot)$ denotes the determinant of a matrix.

\section{System Model}\setlength\arraycolsep{1.5pt}
In this section, we shall introduce the computation model of wireless distributed computing system, followed by proposing a co-channel transmission communication model for data shuffling.

\subsection{Computation Model}
\begin{figure}[tb]
  \centering
  \includegraphics[width=\columnwidth]{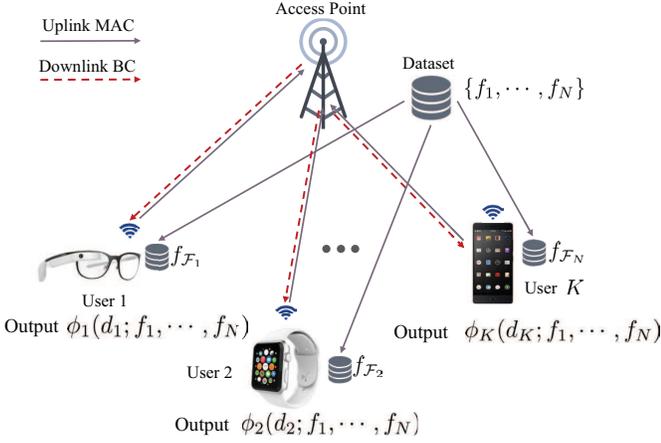}
  \caption{Wireless distributed computing system.}
  \label{fig:wdc}
\end{figure}

Consider the wireless distributed computing system consisting of $K$ mobile users, where mobile users exchange information over a common wirelessly connected access point (AP) as shown in Fig. \ref{fig:wdc}. Suppose each mobile user is equipped with $L$ antennas and the AP uses $M$ antennas. The dataset in the system is assumed to be evenly split to $N$ files $f_1,\cdots,f_N$, each with $F$ bits. Each mobile user $k$ aims to obtain the output of computation task $\phi_k(d_k;f_1,\cdots,f_N)$ with the input $d_k$. For example in object recognition, the dataset is a feature library of various objects. Given the feature vector of an image as input, each mobile user requires the inference result of the image. In practice, the storage size of mobile users is often limited \cite{han2015deep} and the entire dataset cannot be stored directly at the user end. Therefore, we assume that the local memory size of each mobile user is only $\mu F$ bits ($\mu<N$), while the whole dataset can be distributively stored across $K$ mobile users (i.e., $\mu K\geq N$). Let $\mathcal{F}_k\subseteq [N]$ be the index set of files stored at user $k$. Then we have $|\mathcal{F}_k|\leq \mu$ and $\cup_{k\in[K]}\mathcal{F}_k = [N]$. We thus use $f_{\mathcal{F}_k}=\{f_{n}:n\in\mathcal{F}_k\}$ to denote the set of locally stored files at the $k$-th mobile user.

\begin{figure}[tb]
  \centering
  \includegraphics[width=0.9\columnwidth]{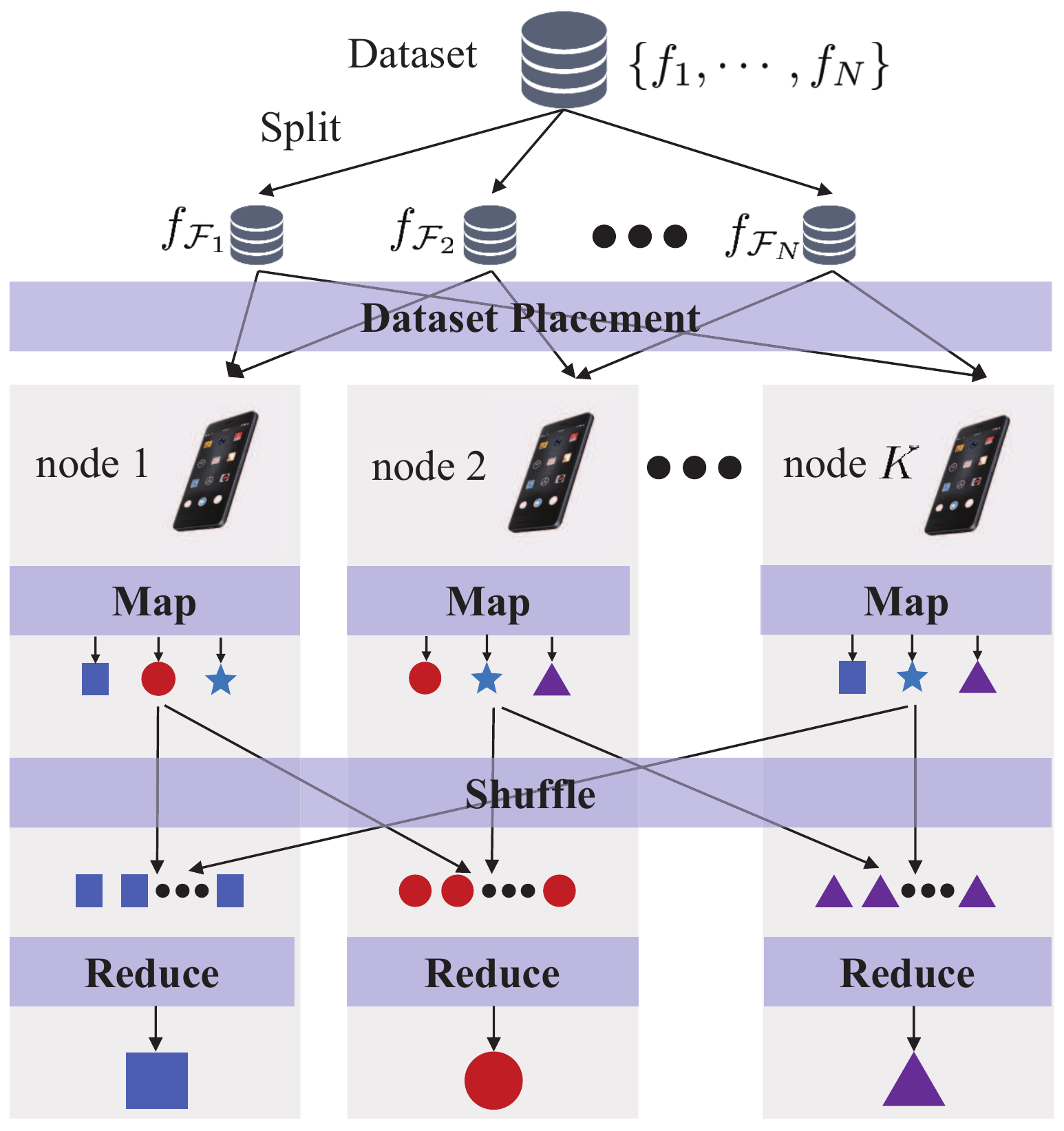}
  \caption{Distributed computing model.}\label{fig:mapreduce}
\end{figure}

In this work, popular distributed computing framework such as MapReduce \cite{dean2008mapreduce} and Spark is adopted to accomplish all computation tasks, where each computation task $\phi_k$ is assumed to be decomposed as \cite{Li2017ascalable}
\begin{equation}\label{eq:mapreduce}
  \phi_k(d_k;f_1,\cdots,f_N)= h_k(g_{k,1}(d_k;f_{1}),\cdots,g_{k,N}(d_k;f_{N})).
\end{equation}
In the focused distributed computing architecture, \textit{Map} function $g_{k,n}(d_k;f_n)$ is computed by the $k$-th mobile user according to file $f_n$, whose output is the intermediate value $w_{k,n}$ with $E$ bits. The \textit{Reduce} function $h_k$ maps all intermediate values $w_{k,1},\cdots,w_{k,N}$ into the output of computation task $\phi_k$. We assume that intermediate values are small enough to be stored at each mobile user while collecting all inputs $d_k$'s has negligible commmunication overhead. As shown in Fig. \ref{fig:mapreduce}, all computation tasks hence can be accomplished via the following four phases:
\begin{itemize}
	\item \textbf{Dataset Placement Phase:} In this phase, the file placement strategy $\mathcal{F}_k$ shall be determined, and files are delivered to the corresponding mobile users in advance to execute Map Phase.
	\item \textbf{Map Phase:} In this phase, intermediate values $w_{k,n}$ are computed locally with map functions $g_{k,n}$ for all $k\in[K]$ and $n\in\mathcal{F}_k$ based on the files $f_{\mathcal{F}_k}$ in the local memory of mobile user $k$.
	\item \textbf{Shuffle Phase:} The output of computation task $\phi_k$ for mobile user $k$ relies on the intermediate values $\{w_{k,n}:n\notin\mathcal{F}_k\}$ that can only be computed by other mobile users in the Map phase. Therefore, mobile users shall exchange intermediate values wirelessly with each other in this phase.
	\item \textbf{Reduce Phase:} By mapping all required intermediate values into the output value, i.e., $\phi_k(d_k;f_1,\cdots,f_N)$ $=h_k(w_{k,1},\cdots,w_{k,N})$, mobile users construct the output of each computation task $\phi_k$.
\end{itemize}
With limited radio resources, data shuffling across mobile devices becomes the significant bottleneck for scaling up wireless distributed computing.

\subsection{Communication Model}
In wireless distributed computing systems, communication often becomes the key bottleneck \cite{Li2017ascalable} \cite{lin2017deep} to accomplish the computation tasks. In this paper, we aim to improve the communication efficiency for the Shuffle Phase given the dataset placement. We shall propose a co-channel transmission framework to efficiently exchange the intermediate values for the data shuffling by modeling this problem as a side information aided message deliveray problem. Specifically, the set of all intermediate values $\{w_{1,1},\cdots,w_{1,N},\cdots,w_{K,N}\}$ is treated as a library of independent messages $\{W_1,\cdots,W_T\}$ with $T=KN$, i.e., the intermediate value $w_{k,n}$ is represented by message $W_{(k-1)N+n}$. Let $\mathcal{T}_k\subseteq[T]$ be the index set of intermediate values available at mobile user $k$, i.e., $\mathcal{T}_k=\{(j-1)N+n:j\in[K],n\in\mathcal{F}_k\} $. Likewise, let $\mathcal{R}_k\subseteq[T]$ be the index set of intermediate values required by mobile user $k$ where $\mathcal{R}_k=\{(k-1)N+n:n\notin\mathcal{F}_k\}$. Note that $\cup_{k\in[K]}\mathcal{T}_k=[T], \mathcal{T}_k\cap \mathcal{R}_k=\emptyset$ due to the structure of MapReduce-like distributed computing framework. With these notations, the data shuffling in Shuffle Phase is modeled as a side information aided message delivery problem. The proposed communication model in Shuffle Phase consists of \textit{uplink multiple access (MAC) stage} and \textit{downlink broadcasting (BC) stage}, as shown in Fig. \ref{fig:wdc}. In uplink MAC stage, the AP collects the mixed signal transmitted by all mobile users, and forwards it to each mobile user in downlink BC stage.

Let the aggregated signal transmitted by mobile user $k$ over $r$ channel uses be
\begin{equation}
  \bs{x}_k=[\bs{x}_k[i]]=\begin{bmatrix}
    \bs{x}_k[1]\\\vdots \\\bs{x}_k[L]
  \end{bmatrix}\in\mathbb{C}^{Lr},
\end{equation}
where $\bs{x}_k[i]\in\mathbb{C}^{r}$ corresponds to the $i$-th antenna. Let $H_k^{\text{up}}[s,i]$ be the channel coefficient between the $i$-th antenna of mobile user $k$ and the $s$-th antenna of AP in uplink MAC stage. The received signal $\bs{y}[s]\in\mathbb{C}^{r}$ at the $s$-th antenna of AP is given by
\begin{equation}
	\bs{y}[s] = \sum_{k=1}^{K}\sum_{i=1}^{L}H_{k}^{\text{up}}[s,i]\bs{x}_k[i]+\bs{n}^{\text{up}}[s],
\end{equation}
where $\bs{n}^{\text{up}}[s]\in\mathbb{C}^{r}$ is the additive isotropic white Gaussian noise. Here, we consider a quasi-static fading channel model in which channel coefficients remain unchanged over $r$ channel uses. By denoting 
\begin{align}
	&\bs{y}=\begin{bmatrix}
		\bs{y}[1] \\ \vdots \\ \bs{y}[M]
	\end{bmatrix}\in\mathbb{C}^{Mr},
	\bs{n}^{\text{up}}=\begin{bmatrix}
		\bs{n}^{\text{up}}[1] \\ \vdots \\ \bs{n}^{\text{up}}[M]
	\end{bmatrix}\in\mathbb{C}^{Mr},\\
	&\bs{H}_{k}^{\text{up}}=\begin{bmatrix}
		H_{k}^{\text{up}}[1,1] & \cdots & H_{k}^{\text{up}}[1,L] \\
		\vdots & \ddots & \vdots \\
		H_{k}^{\text{up}}[M,1] & \cdots & H_{k}^{\text{up}}[M,L]
	\end{bmatrix}\in\mathbb{C}^{M\times L},
\end{align}
the received signal at AP can be written more compactly as
\begin{equation}
	\bs{y} = \sum_{k=1}^{K}(\bs{H}_{k}^{\text{up}}\otimes\bs{I}_r)\bs{x}_k+\bs{n}^{\text{up}},
\end{equation}
where $\otimes$ denotes Kronecker product.

In the downlink BC stage, the AP forwards the received signal $\bs{y}$ to each mobile user. Similarly, the received signal $\bs{z}_k\in\mathbb{C}^{Lr}$ by the $k$-th mobile user is given by
\begin{equation}
	\bs{z}_k = (\bs{H}_{k}^{\text{down}}\otimes\bs{I}_r)\bs{y}+\bs{n}_k^{\text{down}},
\end{equation}
where the channel coefficient matrix $\bs{H}_{k}^{\text{down}}$ in downlink BC stage and the downlink additive isotropic white Gaussian noise $\bs{n}_k^{\text{down}}$ are given as
\begin{align}
	\bs{H}_{k}^{\text{down}}&=\begin{bmatrix}
		H_{k}^{\text{down}}[1,1] & \cdots & H_{k}^{\text{down}}[1,M] \\
		\vdots & \ddots & \vdots \\
		H_{k}^{\text{down}}[L,1] & \cdots & H_{k}^{\text{down}}[L,M]
	\end{bmatrix}\in\mathbb{C}^{L\times M},\\
	\bs{n}_k^{\text{down}}&=\begin{bmatrix}
		\bs{n}_k^{\text{down}}[1] \\ \vdots \\ \bs{n}_k^{\text{down}}[L]
	\end{bmatrix}\in\mathbb{C}^{Lr}.
\end{align}
Therefore, the overall input-output relationship from all mobile users to mobile user $k$ through both the uplink MAC stage and downlink BC stage can be represented as
\begin{align}
		\bs{z}_k &=  \sum_{i=1}^{K}(\bs{H}_{k}^{\text{down}}\otimes\bs{I}_r)(\bs{H}_{i}^{\text{up}}\otimes\bs{I}_r)\bs{x}_i \\
    &\qquad\qquad +(\bs{H}_{k}^{\text{down}}\otimes\bs{I}_r)\bs{n}^{\text{up}}+\bs{n}_k^{\text{down}} \nonumber \\
		&=\sum_{i=1}^{K} (\bs{H}_{ki}\otimes \bs{I}_r)\bs{x}_i+\bs{n}_k,
\end{align}
where $\bs{H}_{ki}=\bs{H}_{k}^{\text{down}}\bs{H}_{i}^{\text{up}}$ denotes the equivalent channel state matrix and $\bs{n}_k=(\bs{H}_{k}^{\text{down}}\otimes\bs{I}_r)\bs{n}^{\text{up}}+\bs{n}_k^{\text{down}}$ is the effective additive noise.

\subsection{Achievable Data Rates and DoF}
Let $R_k(W_l)$ be the achievable data rate of the required message $W_l$ for mobile user $k$. Then there exists certain coding scheme such that the rate of message $W_l$ is $R_k(W_l)$ while the error probability of decoding $W_l$ for mobile user $k$ can be made arbitrarily small as the length of codewords approaches infinity \cite{cover2012elements}. 

As a first-order characterization of channel capacity, degree-of-freedom (DoF) analysis and optimization are widely applied in interference channels \cite{cadambe2008interference,shi2016low,bresler2014feasibility}. The optimal DoF is also charaterized in \cite{cadambe2008interference} for the fully connected $K$ user interference channel.  
Let $\text{SNR}_{k,l}$ be the signal-to-noise-ratio (SNR) therein, followed by the definition of degree-of-freedom \cite{cadambe2008interference}
\begin{equation}
	\textrm{DoF}_{k,l} \stackrel{\triangle}{=} \mathop{\textrm{lim sup}}_{\textrm{SNR}_{k,l}\rightarrow \infty} \frac{R_{k}(W_l)}{\log(\textrm{SNR}_{k,l})}.
\end{equation}
Achievable DoF allocation set is denoted by $\{\textrm{DoF}_{k,l}:k\in[K],l\notin\mathcal{F}_k\}$ and symmetric DoF (denoted by $\textrm{DoF}_{\textrm{sym}}$) is defined as the largest achievable DoF for all $k,l$. That is, the DoF allocation
\begin{equation}
	\{\textrm{DoF}_{k,l}=\textrm{DoF}_{\textrm{sym}}:k\in[K],l\notin\mathcal{F}_k\}
\end{equation}
is achievable. In this paper, we choose DoF as the performance metric for alleviating the interferences in data shuffling. Without loss of generality, we shall maximize the achievable symmetric DoF for the data shuffling in wireless distributed computing, though it can be readily extended to general asymmetric cases.

\section{Interference Alignment Conditions and Low-Rank Framework for Data Shuffling}
In this section, we shall establish the interference alignment conditions for data shuffling in wireless distributed computing, before developing a low-rank optimization framework for the achievable DoF maximization in linear transceiver design. 

\subsection{Interference Alignment Conditions}
Linear coding schemes for transceiver design have found applications in interference alignment \cite{cadambe2008interference} and index coding \cite{maleki2014index} owing to its low-complexity and optimality in terms of DoFs. Therefore, we focus on linear coding scheme in this work. Let $\bs{s}_j\in\mathbb{C}^{d}$ be the representative vector for message $W_j$ with $d$ datastreams such that each datastream carries one DoF. Then the transmitted signal of user $k$ is
\begin{equation}
	\bs{x}_k = \sum_{j\in\mathcal{T}_k}\bs{V}_{kj}\bs{s}_{j},
\end{equation}
where $\bs{V}_{kj}$ is the precoding matrix of mobile user $k$ for message $j$ and formed by
\begin{equation}
	\bs{V}_{kj} = \begin{bmatrix}
		\bs{V}_{kj}[1] \\ \vdots \\ \bs{V}_{kj}[L]
	\end{bmatrix}\in\mathbb{C}^{rL\times d},
\end{equation}
in which $\bs{V}_{kj}[i]\in\mathbb{C}^{r\times d}$ corresponds to the $i$-th antenna of mobile user $k$ over $r$ channel uses. Likewise, let $
	\bs{U}_{kl}= \begin{bmatrix}
		\bs{U}_{kl}[1] & \cdots & \bs{U}_{kl}[L]
	\end{bmatrix}\in\mathbb{C}^{d\times Lr}
$ be the decoding matrix for each message $W_l$ with $l\in\mathcal{R}_k$. We then decode message $W_l$ from
\begin{equation}
	\tilde{\bs{z}}_{kl}=\bs{U}_{kl}\bs{z}_k = \bs{U}_{kl}\sum_{i=1}^{K} (\bs{H}_{ki}\otimes \bs{I}_r)\sum_{j\in\mathcal{T}_i}\bs{V}_{ij}\bs{s}_{j}+\tilde{\bs{n}}_{kl},
\end{equation}
where $\tilde{\bs{n}}_{kl}=\bs{U}_{kl}\bs{n}_k$. We observe that $\tilde{\bs{z}}_{kl}$ contains the linear combination of the entire message set, which can be decomposed into three parts: the desired message, interferences, and locally available messages, i.e.,
\begin{align}
	\tilde{\bs{z}}_{kl}=\mathcal{I}_1(\!\!\!\!\!\!\!\!\underbrace{\bs{s}_l}_{\text{desired message}}\!\!\!\!\!\!\!\!)+&\mathcal{I}_2(\!\!\!\underbrace{\{\bs{s}_j:j\in\mathcal{T}_k\}}_{\text{locally available messages}}\!\!\!) \nonumber\\&+\mathcal{I}_3(\underbrace{\{\bs{s}_j:j\notin \mathcal{T}_k\cup\{l\}\}}_{\text{interferences}})+\tilde{\bs{n}}_{kl}.
\end{align}
Specifically, linear operators $\mathcal{I}_1,\mathcal{I}_2,\mathcal{I}_3$ are given by
\begin{align*}
	\mathcal{I}_1(\bs{s}_l)&=\sum_{i:l\in\mathcal{T}_i}\bs{U}_{kl}(\bs{H}_{ki}\otimes \bs{I}_r)\bs{V}_{il}\bs{s}_l, \\
	\mathcal{I}_2(\{\bs{s}_j:j\in\mathcal{T}_k\})&= \sum_{j\in\mathcal{T}_k}\sum_{i:j\in\mathcal{T}_i}\bs{U}_{kl}(\bs{H}_{ki}\otimes \bs{I}_r)\bs{V}_{ij}\bs{s}_j,\\
	\mathcal{I}_3(\{\bs{s}_j:j\notin \mathcal{T}_k\cup\{l\}\})&=\sum_{j\notin \mathcal{T}_k\cup\{l\}}\sum_{i:j\in\mathcal{T}_i}\bs{U}_{kl}(\bs{H}_{ki}\otimes \bs{I}_r)\bs{V}_{ij}\bs{s}_j.
\end{align*}
Interference alignment \cite{cadambe2008interference} turns out to be a powerful tool to handle the mutual interference among users. The basic idea is to make signals resolvable at intended receivers while aligning and cancelling signals at unintended receivers. To eliminate interferences which is the key limit factor for achieving high data rates, we establish the following interference alignment conditions
\begin{eqnarray}
		\!\!\!\!\!\!\!\!\!\textrm{det}\left(\sum_{i:l\in\mathcal{T}_i}\bs{U}_{kl}(\bs{H}_{ki}\otimes\bs{I}_r)\bs{V}_{il}\right) &\ne& 0, \label{cond:IAC1}\\
		\sum_{i:j\in\mathcal{T}_i}\bs{U}_{kl}(\bs{H}_{ki}\otimes\bs{I}_r)\bs{V}_{ij} &=& \bs{0},~j\notin\mathcal{T}_k\cup\{l\}, \label{cond:IAC2}
\end{eqnarray}
where $l\in\mathcal{R}_k, k\in [K]$. By designing transceivers to satisfy interference alignment conditions (\ref{cond:IAC1}) and (\ref{cond:IAC2}), message $W_l$ can be decoded from signal $\tilde{\bs{s}}_l = \mathcal{I}_1^{-1}\left(\tilde{\bs{z}}_{kl}-\mathcal{I}_2(\{\bs{s}_j:j\in\mathcal{T}_k\})\right)$ for all $l\in\mathcal{R}_k,k\in[K]$.

If conditions (\ref{cond:IAC1}) and (\ref{cond:IAC2}) are met, we can obtain interference-free channels for transmitting $d$-dimensional messages over $r$ channel uses. The achievable $\textrm{DoF}_{k,l}$ is thus given by $d/r$. Hence the symmetric DoF in the wireless distributed computing system is given by
\begin{equation}
		\textrm{DoF}_{\text{sym}}=d/r. \label{eq:dof}
\end{equation}
Consequently, achievable symmetric DoF can be maximized by finding the minimum channel use $r$ subject to (\ref{cond:IAC1}) and (\ref{cond:IAC2}).

\subsection{Low-Rank Optimization Approach}
In this subsection, we develop a low-rank model to establish the interference alignment conditions (\ref{cond:IAC1}) and (\ref{cond:IAC2}) for data shuffling in wireless distributed computing. Note that
\begin{equation}
	\!\!\bs{U}_{kl}(\bs{H}_{ki}\otimes\bs{I}_r)\bs{V}_{ij}=\sum_{m=1}^{L}\sum_{n=1}^{L}H_{ki}[m,n]\bs{U}_{kl}[m]\bs{V}_{ij}[n],
\end{equation}
where $H_{ki}[m,n]$ is the $(m,n)$-th entry of matrix $\bs{H}_{ki}$. 
Define a set of matrices
\begin{align}
	\bs{X}_{k,l,i,j} &= [\bs{X}_{k,l,i,j}[m,n]] = [\bs{U}_{kl}[m]\bs{V}_{ij}[n]]  \\
	&= \begin{bmatrix}
		\bs{U}_{kl}[1]\bs{V}_{ij}[1] & \cdots & \bs{U}_{kl}[1]\bs{V}_{ij}[L] \\
		\vdots & \ddots & \vdots \\
		\bs{U}_{kl}[L]\bs{V}_{ij}[1] & \cdots & \bs{U}_{kl}[L]\bs{V}_{ij}[L]
	\end{bmatrix} \\
	&= \begin{bmatrix}
		\bs{U}_{kl}[1] \\
		\vdots \\
		\bs{U}_{kl}[L]
	\end{bmatrix}\begin{bmatrix}
		\bs{V}_{ij}[1] & \cdots & \bs{V}_{ij}[L]
	\end{bmatrix}\\
	&=\tilde{\bs{U}}_{kl}\tilde{\bs{V}}_{ij},
\end{align}
where $\tilde{\bs{U}}_{kl}\in\mathbb{C}^{Ld \times r}$ and $\tilde{\bs{V}}_{ij}\in\mathbb{C}^{r\times Ld}$. We further denote
\begin{align}
	\bs{X}&= [\bs{X}_{k,l,i,j}] \\
		&= \begin{bmatrix}
		\bs{X}_{1,1,1,1} & \cdots &\bs{X}_{1,1,1,T} & \cdots &\bs{X}_{1,1,K,T} \\
		\vdots & \ddots & \vdots & \ddots & \vdots\\
		\bs{X}_{1,T,1,1} & \cdots &\bs{X}_{1,T,1,T} & \cdots &\bs{X}_{1,T,K,T} \\
		\vdots & \ddots & \vdots & \ddots & \vdots\\
		\bs{X}_{K,T,1,1} & \cdots &\bs{X}_{K,T,1,T} &\cdots &\bs{X}_{K,T,K,T}
	\end{bmatrix} \\
	&=\begin{bmatrix}
		\tilde{\bs{U}}_{11} \\
		\vdots \\
		\tilde{\bs{U}}_{1T} \\
		\vdots \\
		\tilde{\bs{U}}_{KT}
	\end{bmatrix}\begin{bmatrix}
		\tilde{\bs{V}}_{11} & \cdots & \tilde{\bs{V}}_{1T} & \cdots & \tilde{\bs{V}}_{KT}
	\end{bmatrix} \\
	&=\tilde{\bs{U}}\tilde{\bs{V}},
\end{align}
where $\tilde{\bs{U}}\in\mathbb{C}^{LdKT \times r}$ and $\tilde{\bs{V}}\in\mathbb{C}^{r\times LdKT}$. Without loss of generality, to enable efficient algorithms design, we set $\sum_{i:l\in\mathcal{T}_i}\bs{U}_{kl}(\bs{H}_{ki}\otimes\bs{I}_r)\bs{V}_{il}=\bs{I}$ in (\ref{cond:IAC1}). Then the interference alignment conditions (\ref{cond:IAC1}) and (\ref{cond:IAC2}) can be rewritten as 
\begin{eqnarray}
\!\!\!\!\!\!\sum_{i:l\in\mathcal{T}_i}\sum_{m=1}^{L}\sum_{n=1}^{L}H_{ki}[m,n]\bs{X}_{k,l,i,l}[m,n]\!\!\!&=&\!\!\!\bs{I}, \label{cond:IAC_af1}\\
\!\!\!\!\!\!\sum_{i:j\in\mathcal{T}_i}\sum_{m=1}^{L}\sum_{n=1}^{L}H_{ki}[m,n]\bs{X}_{k,l,i,j}[m,n]\!\!\!&=&\!\!\! \bs{0},~j\notin\mathcal{T}_k\cup\{l\}, \label{cond:IAC_af2}
\end{eqnarray}
which can be charaterized by $\mathcal{A}(\bs{X})=\bs{b}$ with the linear operator $\mathcal{A}:\mathbb{C}^{D\times D}\mapsto \mathbb{C}^{S}$ as a function of $\{\bs{H}_{ki}\}$. 
Note that the rank of matrix $\bs{X}$ is equal to the number of channel uses $r$ since $\bs{X}=\tilde{\bs{U}}\tilde{\bs{V}}$, i.e.,
\begin{equation}
	\textrm{rank}(\bs{X})=r.
\end{equation}
We hence propose the following low-rank optimization approach to maximize the achievable symmetric DoF
\begin{eqnarray}
		\mathscr{P}:\mathop{\textrm{minimize}}_{\bs{X}\in\mathbb{C}^{D\times D}} && \textrm{rank}(\bs{X}) \nonumber \\
		\textrm{subject to} && \mathcal{A}(\bs{X})=\bs{b},
\end{eqnarray}
where $D=LdKT$. However, problem $\mathscr{P}$ is computationally hard due to the non-convexity of the rank function. 

\subsection{Problem Analysis}
Low-rank optimization approach has recently caught enormous attentions particularly in machine learning, high-dimensional statistics, and recommendation systems \cite{davenport2016overview}. Unfortunately, low-rank optimization problems are generally intractable due to the non-convex rank function. Therefore, many efforts focused on finding tractable representation for the rank function, based on which a number of algorithms are developed. 

\subsubsection{Nuclear Norm Relaxation}\label{subsubsection:nuclearnorm}
Nuclear norm \cite{davenport2016overview} has demonstrated its effectiveness as the convex surrogate for the rank function, yielding the following nuclear norm minimization problem
\begin{eqnarray}
\mathop{\textrm{minimize}}_{{\bs{X}}}&& \|{\bs{X}}\|_* \nonumber\\
\textrm{subject to}&& \mathcal{A}({\bs{X}})={\bs{b}}.\label{prob:nuc}
\end{eqnarray}
The nuclear norm $\|\bs{X}\|_*$ is equal to the sum of the
singular values of $\bs{X}$. It is the convex hull of the collection of atomic unit-norm rank-one matrices, and is thus the tightest convex relaxation of the rank function. Its equivalent semidefinite programming (SDP) form 
\begin{eqnarray}
\mathop{\textrm{minimize}}_{\bs{X},\bs{W}_1,\bs{W}_2}&& \textrm{Tr}(\bs{W}_1)+\textrm{Tr}(\bs{W}_2) \nonumber\\
\textrm{subject to}&& \mathcal{A}({\bs{X}})={\bs{b}}, \label{prob:nuc} \\
&& \begin{bmatrix}
  \bs{W}_1 & \bs{X} \\ \bs{X}^{\sf{H}} & \bs{W}_2
\end{bmatrix}\succeq \bs{0} \nonumber
\end{eqnarray}
can be solved by the interior point method with high precision at a low iteration count. However, this second-order algorithm has high computational complexity with computational cost $\mathcal{O}((S+D^2)^3)$ at each iteration due to the Newton step \cite{boyd2004convex}. The first-order alternating direction method of multipliers (ADMM) \cite{shi2015large_ADMM,o2016conic} significantly reduces the computational cost to $\mathcal{O}(SD^2+D^3)$ for each iteration (please refer to \ref{subsubsection:complexity} for more details). It converges within $\mathcal{O}(1/\epsilon)$ iterations given the precision $\epsilon>0$. 

However, the nuclear norm minimization approach yields unsatisfactory performance due to the poor structure of the affine constraint in problem $\mathscr{P}$. For example, in the scenario of two users with $K=N=2,\mu=d=L=M=1$, each mobile user stores distinct files locally, and requires the intermediate values computed by the other one. In this case, problem $\mathscr{P}$ is

\begin{eqnarray}
\mathop{\textrm{minimize}}_{\bs{X}}&&\textrm{rank}(\bs{X})\nonumber \\
 \textrm{subject to}&&\bs{X}=\left[\begin{array}{*{9}{@{}C{\mycolwd}@{}}}
 \star &\star &\star &\star &\star &\star &\star &\star \\
  \star &\star &\star &\star &\star &\star &\star &\star \\
    \star &\star &\star &\star &   \star & \star & \frac{1}{H_{12}} & 0 \\
    \star &\star &\star &\star &\star &\star &\star &\star \\
    \star &\star &\star &\star &\star &\star &\star &\star \\
        0 & \frac{1}{H_{21}}& \star & \star &\star &\star &\star &\star \\
        \star &\star &\star &\star &\star &\star &\star &\star \\
        \star &\star &\star &\star &\star &\star &\star &\star 
      \end{array}\right],
\end{eqnarray}
where the value of $\star$ is unconstrained. In this case, the nuclear norm approach always returns full rank solution while the optimal rank is 1. Furthermore, the numerical results provided in Section V shall demonstrate that the convex relaxation approach yields poor performance on average.

\subsubsection{Schatten-$p$ Norm Approximation and Iterative Reweighted Least Squares Minimization}\label{subsubsection:IRLS}
To provide better approximation for the rank function, Schatten-$p$ norm ($0\leq p\leq 1$)  of a matrix has been studied in \cite{mohan2012iterative}. Specifically, the Schatten-$p$ norm of matrix $\bs{X}\in\mathbb{C}^{D\times D}$ is defined as
\begin{equation}
  \|\bs{X}\|_p = \left(\sum_{i=1}^{D} \sigma_i^p(\bs{X})\right)^{1/p}.
\end{equation}
Since it is nonconvex for $p<1$, an iterative reweighted least squares algorithm (IRLS-$p$) is proposed to alternatively minimize weighted Frobenius norm and update weights $\bs{W}$ based on the observation that
\begin{equation}
  \|\bs{X}\|_p^p=\textrm{Tr}((\bs{X}^{\sf{H}}\bs{X})^{\frac{p}{2}-1}\bs{X}^{\sf{H}}\bs{X})
\end{equation}
holds for non-singular matrix $\bs{X}$. In the $t$-th iteration, $\bs{X}$ and weight matrix $\bs{W}$ can be updated as follows
\begin{align}
   \bs{X}^{[t]} &= \mathop{\textrm{argmin}}_{\bs{X}} \{\textrm{Tr}(\bs{W}^{[k-1]}\bs{X}^{\sf{H}}\bs{X}):\mathcal{A}({\bs{X}})={\bs{b}}\} \label{iter:IRLS1}\\
   \bs{W}^{[t]}&= ({\bs{X}^{[t]}}^{\sf{H}}\bs{X}^{[t]}+\gamma^{[k]}\bs{I})^{\frac{p}{2}-1},\label{iter:IRLS2}
 \end{align} 
where $\gamma^{[k]}\in\mathbb{R}$ is a regularization parameter to ensure that $\bs{W}^{[t]}$ is well-defined and $\{\gamma^{[k]}\}$ is a non-increasing sequence. However, its performance still falls short when applied to problem $\mathscr{P}$ given the poorly structured affine constraint. In this work, we shall propose a novel difference-of-convex-functions (DC) algorithm to achieve considerable performance improvements by rewriting the rank function as a DC function.

\section{DC Approach for Low-Rank Optimization}
This section develops a DC algorithm for the low-rank optimization problem in data shuffling. This is achieved by proposing a novel DC representation for the rank function, and developing an efficient DC algorithm based on the proposed DC representation.

\subsection{DC Approach}
A DC representation of the rank function has recently been proposed in \cite{gotoh2017dc}, followed by a DC algorithm to solve problem $\mathscr{P}$. We will first introduce the definition of Ky Fan norm.
\begin{definition}{Ky Fan $k$-norm \cite{watson1993matrix}:}\label{def:ky} The Ky Fan $k$-norm of a matrix $\bs{X}$ is a convex function of matrix $\bs{X}$ and given by the sum of its largest-$k$ singular values, i.e.,
\begin{equation}
  |\!|\!|\bs{X}|\!|\!|_k=\sum_{i=1}^{k}\sigma_i(\bs{X}),
\end{equation}
where $\sigma_i(\bs{X})$ is the $i$-th largest singular value of $\bs{X}$. 
\end{definition}
Based on Definition \ref{def:ky}, if a matrix is low-rank (rank $r$), its Ky Fan $r$-norm equals its nuclear norm. Then a DC representation for the rank function can be obtained. For any matrix $\bs{X}\in\mathbb{C}^{m\times n}$, the following equation holds \cite{gotoh2017dc}:
\begin{equation}
  \textrm{rank}(\bs{X})=\min\{k:\|\bs{X}\|_*-|\!|\!|\bs{X}|\!|\!|_k=0,k\leq \min\{m,n\}\}.
\end{equation}

Therefore, by representing the rank function with Ky Fan $k$-norm, problem $\mathscr{P}$ can be solved by finding the minimum $k$ such that the optimal objective value is zero in the following optimization problem:
\begin{eqnarray}
    \mathop{\textrm{minimize}}_{\bs{X}\in\mathbb{C}^{D\times D}} && \|\bs{X}\|_*-|\!|\!|\bs{X}|\!|\!|_k \nonumber \\
    \textrm{subject to} && \mathcal{A}(\bs{X})=\bs{b}, \label{prob:dc_nuc}
\end{eqnarray}
where the objective is the difference of two convex functions $\|\bm{X}\|_*$ and $|\!|\!|\bs{X}|\!|\!|_k$. Due to the nonconvex DC objective function, the majorization-minimization (MM) algorithm \cite{tao1997convex,le2018dc} can be adopted to iteratively solve a convex subproblem by linearizing $|\!|\!|\bs{X}|\!|\!|_k$ as $\textrm{Tr}(\partial|\!|\!|\bs{X}_{t}|\!|\!|_k^{\sf{H}}\bs{X})$, i.e., by solving
\begin{eqnarray}
    \mathop{\textrm{minimize}}_{\bs{X}\in\mathbb{C}^{D\times D}} && \|\bs{X}\|_*-\textrm{Tr}(\partial|\!|\!|\bs{X}_{t}|\!|\!|_k^{\sf{H}}\bs{X}) \nonumber \\
    \textrm{subject to} && \mathcal{A}(\bs{X})=\bs{b} \label{prob:subDCA}
\end{eqnarray}
in the $(t+1)$-th iteration. Here $\bs{X}_{t}$ is the solution to (\ref{prob:subDCA}) in the $t$-th iteration. $\partial|\!|\!|\bs{X}_t|\!|\!|_k$ \cite{watson1993matrix} denotes the subdifferential of $|\!|\!|\bs{X}|\!|\!|_k$ at $\bs{X}_t$ and can be chosen as
\begin{equation}
  \partial|\!|\!|\bs{X}_t|\!|\!|_k = \{\bs{U}\textrm{diag}(\bs{q})\bs{V}^{\sf{H}},\bs{q}=[\underbrace{1,\cdots,1}_{k},\underbrace{0,\cdots,0}_{D-k}]\}, \label{subg:DCnuc}
\end{equation} 
where $\bs{X}_t=\bs{U}\bs{\Sigma}\bs{V}^{\sf{H}}$ is the singular value decomposition (SVD) of $\bs{X}_t$. 

Unfortunately, the main drawback of this DC approach is that in each iteration a nuclear norm minimization problem (\ref{prob:subDCA}) should be solved. The computational cost of nuclear norm minimization problem is $\mathcal{O}(\frac{1}{\epsilon}(SD^2+D^3))$ even with first-order ADMM algorithm for precision $\epsilon$, which is computationally costly and not amenable to the data shuffling problem in this paper. Efficient algorithm should be proposed especially for the wireless distributed computing scenarios with large number of mobile users. Next, we shall propose a novel computationally efficient DC approach for solving problem $\mathscr{P}$, for which we propose a novel DC representation for the rank function.

\subsection{A Novel DC Representation for Rank Function}
We observe that the nuclear norm function in the objective function of problem (\ref{prob:dc_nuc}) leads to cumbersome computations. To overcome the drawback, we propose a novel DC representation of the rank function. We first introduce:
\begin{definition}
	For any integer $1\leq k\leq \min\{m,n\}$, the Ky Fan $2$-$k$ norm \cite{doan2016finding} of matrix $\bs{X}\in\mathbb{C}^{m\times n}$ is defined as the $\ell_2$-norm of the subvector formed by the largest-$k$ singular values of $\bs{X}$. That is,  
	\begin{equation}
		|\!|\!|\bs{X}|\!|\!|_{k,2} = \left(\sum_{i=1}^{k}\sigma_i^2(\bs{X})\right)^{1/2},
	\end{equation}
	where $\sigma_i(\bs{X})$ is the $i$-th largest singular value of matrix $\bs{X}$.
\end{definition}
The Ky Fan $2$-$k$ norm is a unitarily invariant norm, and can be computed via the following SDP problem \cite{doan2016finding}
\begin{eqnarray}
    |\!|\!|\bs{X}|\!|\!|_{k,2}^2 = \mathop{\textrm{minimize}}_{z,\bs{U}} && kz+\textrm{Tr}(\bs{U}) \nonumber \\
    \textrm{subject to} && z\bs{I}+\bs{U} \succeq \bs{X}^{\sf{H}}\bs{X},\\
    && \bs{U}\succeq\bs{0}. \nonumber
\end{eqnarray}
Note that $\textrm{rank}(\bs{X})= r$ means that the $\min\{m,n\}-r$ smallest singular values of matrix $\bs{X}\in\mathbb{C}^{m\times n}$ are zeros. Based on this fact, we have the following proposition:
\begin{proposition}
	For a matrix $\bs{X}\in\mathbb{C}^{m\times n}$, we have
	\begin{equation}
		\textrm{rank}(\bs{X})\leq k \Leftrightarrow \|\bs{X}\|_F = |\!|\!|\bs{X}|\!|\!|_{k,2}.
	\end{equation}	
	Futhermore,
	\begin{equation}
		\textrm{rank}(\bs{X})=\min\{k:\|\bs{X}\|_F^2-|\!|\!|\bs{X}|\!|\!|_{k,2}^2=0,k\leq\min\{m,n\}\}.
	\end{equation}
\end{proposition}
\begin{proof}
	Given $\textrm{rank}(\bs{X})\leq k$, we have $\sigma_i(\bs{X})=0~\forall i>k$. It  follows that $\|\bs{X}\|_F=|\!|\!|\bs{X}|\!|\!|_{k,2}$. Conversely, we can deduce $\sigma_i(\bs{X})=0~\forall i>k$ from $\|\bs{X}\|_F=|\!|\!|\bs{X}|\!|\!|_{k,2}$. Thus, the rank of matrix $\bs{X}$ is no more than $k$.

	Let the rank of matrix $\bs{X}$ be $r$. Then $\sigma_i(\bs{X})=0~\forall i>r$ and $\sigma_i(\bs{X})>0~\forall i\leq r$. Since $\|\bs{X}\|_F=|\!|\!|\bs{X}|\!|\!|_{k,2}$ if and only if $\textrm{rank}(\bs{X})\leq k$, the minimum $k$ for $\|\bs{X}\|_F^2 - |\!|\!|\bs{X}|\!|\!|_{k,2}^2=0$ will be exactly $r$. Conversely, $ r=\min\{k:\|\bs{X}\|_F^2 - |\!|\!|\bs{X}|\!|\!|_{k,2}^2=0\}$ we deduce that $\sigma_i(\bs{X})=0~\forall i>r$ and $\sigma_i(\bs{X})>0~\forall i\leq r$. Then $\textrm{rank}(\bs{X})=r$.
\end{proof}

\subsection{Efficient DC Algorithm for Problem $\mathscr{P}$}
With the proposed novel DC representation of rank function, the minimum rank $r$ can be found by sequentially solving 
\begin{eqnarray}
		\mathscr{P}_{\text{DC}}:\mathop{\textrm{minimize}}_{\bs{X}\in\mathbb{C}^{D\times D}} && \|\bs{X}\|_F^2-|\!|\!|\bs{X}|\!|\!|_{k,2}^2 \nonumber \\
		\textrm{subject to} && \mathcal{A}(\bs{X})=\bs{b} \label{prob:dc_fro}
\end{eqnarray}
and incrementing $k$ from $1$ to $\min\{m,n\}$, until the objective value of problem $\mathscr{P}_{\text{DC}}$ achieves zero. Problem $\mathscr{P}_{\text{DC}}$ is a DC programming problem since its objective function is the difference of two convex functions. 

To develop the simplified form of DC algorithm \cite{tao1997convex} , we equivalently
rewrite problem $\mathscr{P}_{\text{DC}}$ as
\begin{eqnarray}
	\mathop{\textrm{minimize}}_{\mathcal{X}\in\mathbb{C}^{m\times n}} && \|\bs{X}\|_F^2+I_{(\mathcal{A}(\bs{X})=\bs{b})}(\bs{X})-|\!|\!|\bs{X}|\!|\!|_{k,2}^2
\end{eqnarray}
where the indicator function $I$ is given by
\begin{equation}
	I_{(\mathcal{A}(\bs{X})=\bs{b})}(\bs{X}) = \left\{\begin{aligned}
		& 0, && \mathcal{A}(\bs{X})=\bs{b} \\
		& +\infty, && \text{otherwise}
	\end{aligned}\right..
\end{equation}
To deal with the complex domain, we employ Wirtinger's calculus \cite{bouboulis2012adaptive}. Let $g(\bs{X})=\|\bs{X}\|_F^2+I_{(\mathcal{A}(\bs{X})=\bs{b})}(\bs{X}), h(\bs{X})=|\!|\!|\bs{X}|\!|\!|_{k,2}^2$. Since $\{\bs{X}:\mathcal{A}(\bs{X})=\bs{b}\}$ is an affine subspace, function $g$ and function $h$ are both convex. We denote
\begin{eqnarray}
	\alpha = \inf_{\bs{X}\in\mathcal{X}} && f(\bs{X}) = g(\bs{X})-h(\bs{X})
\end{eqnarray}
where $\mathcal{X}=\mathbb{C}^{m\times n}$. According to the Fenchel's duality \cite{rockafellar2015convex}, its dual problem is given by
\begin{eqnarray}
	\alpha = \inf_{\bs{Y}\in\mathcal{Y}} && h^*(\bs{Y})-g^*(\bs{Y}).
\end{eqnarray}
Here $h^*(\bs{Y})$ and $g^*(\bs{Y})$ are the conjugate functions of $g$ and $h$ respectively. The conjugate function is defined by
\begin{eqnarray}
	g^*(\bs{Y})=\sup_{\bs{X}\in\mathcal{X}} && \langle \bs{X},\bs{Y} \rangle-g(\bs{X}),
\end{eqnarray}
where the inner product is defined as $\langle \bs{X},\bs{Y} \rangle=\textrm{Tr}(\bs{X}^{\sf{H}}\bs{Y})$ based on \cite{bouboulis2012adaptive}. 

Simplified DC algorithm aims to update both the primal and dual variables via successive convex approximation. Specific iterations for solving problem $\mathscr{P}_{\text{DC}}$ are given by
\begin{align}
	&\bs{Y}^{[t]}=\arg\inf_{\bs{Y}\in\mathcal{Y}}~h^*(\bs{Y})-[g^*(\bs{Y}^{[t-1]})+\langle \bs{Y}-\bs{Y}^{[t-1]}, \bs{X}^{[t]}\rangle],\label{dc:iter1}\\
	&\bs{X}^{[t+1]}=\arg\inf_{\bs{X}\in\mathcal{X}}~g(\bs{X})-[h(\bs{X}^{[t]})+\langle \bs{X}-\bs{X}^{[t]}, \bs{Y}^{[t]}\rangle]. \label{dc:iter2}
\end{align}
Using the Fenchel biconjugation theorem \cite{rockafellar2015convex}, equation (\ref{dc:iter1}) can be summarized as
\begin{equation}
 	\bs{Y}^{[t]} \in \partial h(\bs{X}^{[t]}).
 \end{equation} 
 Therefore, we propose to solve problem $\mathscr{P}_{\text{DC}}$ by updating the primal and dual variables $\bs{X}^{[t+1]},\bs{Y}^{[t]}$ via
\begin{align}
	&\bs{Y}^{[t]} \in \partial |\!|\!|\bs{X}^{[t]}|\!|\!|_{k,2}^2\label{dc:iter3}\\
	&\bs{X}^{[t+1]}=\arg\inf_{\bs{X}\in\mathcal{X}}~\{\|\bs{X}\|_F^2-\langle \bs{X}, \bs{Y}^{[t]}\rangle:\mathcal{A}(\bs{X})=\bs{b}\}.\label{dc:iter4}
\end{align}

\begin{proposition}
	One subgradient of $|\!|\!|\bs{X}|\!|\!|_{k,2}^2$ is given by
	\begin{equation}
		\partial |\!|\!|\bs{X}|\!|\!|_{k,2}^2:=2\bs{U}\bs{\Sigma}_k\bs{V}^{\sf{H}}, \label{eq:subdiff}
	\end{equation}
	where $\bs{X}=\bs{U}\bs{\Sigma}\bs{V}^{\sf{H}}$ is the singular value decomposition (SVD) of matrix $\bs{X}\in\mathbb{C}^{D\times D}$ and $\bs{\Sigma}_k$ keeps the largest $k$ diagonal elements of the matrix $\bs{\Sigma}$.
\end{proposition}
\begin{proof}
	First we note that the Ky Fan $2$-$k$ norm of matrix $\bs{X}$ is orthogonally invariant. This can be obtained from the orthogonal invariance of singular values, and
	\begin{equation}
		|\!|\!|\bs{X}|\!|\!|_{k,2}^2 = |\!|\!|\bs{\sigma}(\bs{X})|\!|\!|_{k,2}^2=\sum_{i=1}^{k}\sigma_i^2(\bs{X}).
	\end{equation}
	Here $\bs{\sigma}=[\sigma_i(\bs{X})]\in\mathbb{R}^{D}$ denotes the vector composed by all singular values of matrix $\bs{X}$. $|\!|\!|\bs{\sigma}(\bs{X})|\!|\!|_{k,2}$ denotes the Ky Fan $2$-$k$ norm of vector $\bs{\sigma}(\bs{X})$. The subgradient of $|\!|\!|\bs{\sigma}(\bs{X})|\!|\!|_{k,2}^2$ with respect to $\bs{\sigma}(\bs{X})$ is given by
	\begin{equation}
		 \bs{c}\in\mathbb{R}^{D}:c_i = \left\{\begin{aligned}
		 	& 2\sigma_i(\bs{X}), && i<=k \\
		 	& 0, && i>k
		 \end{aligned}\right..
	\end{equation}
	According to the subdifferential of orthogonally invariant norm \cite{watson1992characterization}, we obtain
	\begin{equation}
		\{\bs{U}\textrm{diag}(\bs{d})\bs{V}^{\sf{H}}:\bs{X}=\bs{U}\bs{\Sigma}\bs{V}^{\sf{H}},\bs{d}\in\partial |\!|\!|\bs{\sigma}(\bs{X})|\!|\!|_{k,2}\} \subseteq \partial |\!|\!|\bs{X}|\!|\!|_{k,2}.
	\end{equation}
	It then follows that
	\begin{equation}
		2\bs{U}\bs{\Sigma}_k\bs{V}^{\sf{H}} \in \partial |\!|\!|\bs{X}|\!|\!|_{k,2}^2,
	\end{equation}
	where $\bs{\Sigma}_k$ is given by
	\begin{equation}
		(i,j)\text{-th entry of}~\bs{\Sigma}_k:=\left\{\begin{aligned}
			&\sigma_i(\bs{X}), && i=j,i<=k \\
			&0, && \text{otherwise}
		\end{aligned}\right..
	\end{equation}
\end{proof}
Note that each iteration of (\ref{dc:iter3}) and (\ref{dc:iter4}) for the proposed DC algorithm can be computed much more efficiently than solving the nuclear norm minimization problem (\ref{prob:subDCA}) since (\ref{dc:iter4}) is a simple quadratic programming (QP) problem with closed form solutions. Specifically, according to (\ref{dc:iter3}) and (\ref{dc:iter4}), $\bs{X}^{[t+1]}$ can be rewritten as the solution to the following quadratic program:
\begin{eqnarray}
		\mathop{\textrm{minimize}}_{\bs{X}\in\mathbb{C}^{D\times D}} && \|\bs{X}-\frac{1}{2}\partial|\!|\!|\bs{X}^{[t]}|\!|\!|_{k,2}^2\|_F^2 \nonumber \\
		\textrm{subject to} && \mathcal{A}(\bs{X})=\bs{b}.
\end{eqnarray}
The solution to this least square problem with affine constraint is the orthogonal projection onto the affine subspace, whose closed-form is given by
\begin{equation}
	\bs{X}^{[t+1]}=(\bs{I}-\mathcal{A}^+\mathcal{A})(\frac{1}{2}\partial|\!|\!|\bs{X}^{[t]}|\!|\!|_{k,2}^2)+\mathcal{A}^+(\bs{b}),\label{dc:iter}
\end{equation}
where $\mathcal{A}^+=\mathcal{A}^{\sf{H}}(\mathcal{A}\mathcal{A}^{\sf{H}})^{-1}$. Therefore, the overall procedure of our proposed DC algorithm is shown in Algorithm \ref{algorithm:Proposed_DC}. 
 \SetNlSty{textbf}{}{:}
\IncMargin{1em}
\begin{algorithm}[h]
\textbf{Input:} $\mathcal{A},\bm{b}$.\\
 \For{$r=1,\cdots,\min\{m,n\}$}{
 Initialize: $\bm{X}_{r}^{[0]}\in\mathbb{C}_*^{m\times n}$ \\
 \While{not converge}{
 $\bs{X}_r^{[t+1]}=(\bs{I}-\mathcal{A}^+\mathcal{A})(\frac{1}{2}\partial |\!|\!|\bs{X}_r^{[t]}|\!|\!|_{k,2}^2)+\mathcal{A}^+(\bs{b})$}
 \If{$\textrm{rank}(\bs{X}_r)\leq r$}{\Return $\bs{X}_r$}
 }
 \textbf{Output:} $\bs{X}_r$ and $\textrm{rank}(\bs{X}_r)$.
 \caption{Proposed DC Approach for problem $\mathscr{P}$}
 \label{algorithm:Proposed_DC}
\end{algorithm}

\subsection{Computational Complexity and Convergence Analysis}\label{subsubsection:complexity}
The proposed DC algorithm involves computing a series of equation (\ref{dc:iter}) multiple times for fixed rank $r$. Since both $\mathcal{A}$ and $\mathcal{A}^+$ can be computed and stored in advance, in each iteration the computational overhead comes from matrix vector multiplication and subgradient evaluation. Since the dimension of $\mathcal{A}$ is $\mathbb{C}^{D\times D}\mapsto \mathbb{C}^{S}$, the complexity of matrix vector multiplication is $\mathcal{O}(SD^2)$. Computing the subgradient by following (\ref{eq:subdiff}) is dominated by the SVD with computational complexity $\mathcal{O}(D^3)$. Therefore, the computational overhead of the proposed DC algorithm for each iteration is $\mathcal{O}(SD^2+D^3)$. However, the first-order algorithm ADMM \cite{o2016conic} needs to solve a sequence of semidefinite cone projection problem via SVD for solving the nuclear norm minimization problem (\ref{prob:subDCA}), which yields computational cost $\mathcal{O}(\frac{1}{\epsilon}(SD^2+D^3))$ with $\epsilon$ as the solution accuracy. Therefore, our proposed DC algorithm is much more computationally efficient with closed form solution for solving the DC program (\ref{prob:dc_fro}), instead of solving a nuclear norm minimization problem for solving the DC program (\ref{prob:dc_nuc}) using the algorithm in \cite{gotoh2017dc}. The complexity of the iterations (\ref{iter:IRLS1}) and (\ref{iter:IRLS2}) for the IRLS-$p$ algorithm using projected gradient descent method \cite{mohan2012iterative} is $\mathcal{O}((SD^2+D^3)\log\frac{1}{\epsilon})$.

The proposed DC algorithm can be implemented very efficiently due to the sparse structure of operator $\mathcal{A}$. Therefore, the overhead of matrix vector multiplication is often small especially when $L$ and $d$ are much smaller compared with the number of involved mobile users. Specifically, the sparsity level of the linear operator $\mathcal{A}$ is given as
\begin{equation}
\sum_{k=1}^{K}\sum_{l\in\mathcal{R}_k}\sum_{j\ne \mathcal{T}_k}|\{i:j\in \mathcal{T}_i\}|L^2d^2.
\end{equation}
For example, for a single-antenna wireless distributed computing system with $5$ mobile users and $10$ files in the dataset, if each mobile user stores $6$ files in its local storage unit and messages are delivered with single datastream, $D=250$ and the sparisity level of $\mathcal{A}$ is only $920$.

The convergence of the proposed DC algorithm for solving problem $\mathscr{P}_{\text{DC}}$ is given by the following proposition.
\begin{proposition}\label{prop:convergence}
	Given rank parameter $k$, the proposed Algorithm \ref{algorithm:Proposed_DC} for solving problem $\mathscr{P}_{\text{DC}}$ converges to critical points from arbitrary initial points.
\end{proposition}
\begin{proof}
	Please refer to Appendix \ref{append:convergence} for details.
\end{proof}

\section{Numerical Results}
In this section, we describe numerical experiments to compare the performance of the proposed DC algorithm (Algorithm \ref{algorithm:Proposed_DC}) with the following benchmarks:
\begin{itemize}
 	\item \textbf{Nuclear norm relaxation}: To evaluate the performance of the nuclear norm relaxation approach (\ref{prob:nuc}), we implement the interior point method introduced in Section \ref{subsubsection:nuclearnorm} with CVX \cite{cvx} toolbox.
 	\item \textbf{Iterative reweighted least squares (IRLS)}: In \cite{mohan2012iterative}, smoothed Schatten-$p$ norm approximation for the rank function is adopted. To solve this nonconvex problem, the iterative reweighted least squares algorithm is proposed as presented in Section \ref{subsubsection:IRLS}. $p$ is chosen as $0.5$ through cross validation in this section.
 \end{itemize}

In all simulations, we consider the symmetric case where all mobile users and the AP are equipped with $L=M$ antennas. The maximum achievable symmetric DoF (\ref{eq:dof}) is chosen as the performance metric. The channel coefficients are randomly drawn from independent and identically distributed complex Gaussian distribution, i.e., $\bs{H}_{ki}\sim\mathcal{CN}(\bs{0},\bs{I})$. For each algorithm, the rank is determined by the number of singular values above $10^{-5}$. Given $r$, iterations for the proposed DC algorithm will be terminated when the $(r+1)$-th singular value is less than $10^{-5}$, i.e., $\sigma_{r+1}(\bs{X})<10^{-5}$.

\subsection{Achievable DoF over Local Storage Size}
Consider a wireless distributed computing framework with $5$ single-antenna mobile users and a single-antenna AP. Each mobile user stores $5$ to $9$ files locally while the full dataset consists of $10$ files. We shall evaluate the maximum achievable symmetric DoF that each algorithm can obtain with the assumption that each message is a single datastream. We run each algorithm $100$ replications to evaluate the relationship between DoF and the local storage size.

\begin{figure}[h]
  \centering
  \includegraphics[width=\columnwidth]{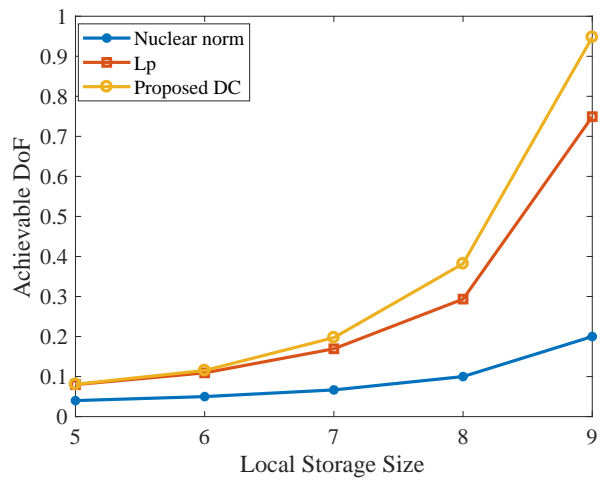}
  \caption{The maximum achievable symmetric DoF over local storage size $\mu$ of each mobile user.}
  \label{fig:cache}
\end{figure}

From Fig. \ref{fig:cache}, we observe that the achievable symmetric DoF has visible growth when more files are stored at each mobile devices for all algorithms. Clearly, this is because more cooperation is enabled and fewer intermediate values need to be exchanged when each mobile user can access more files of the whole dataset. The proposed DC algorithm outperforms both the IRLS algorithm and nuclear norm relaxation. The result of this experiment demonstrates that the proposed DC representation for the rank function has advantages over the Schatten-$p$ norm approximation approach, while the nuclear norm relaxation is inferior to the other two approaches. 

\subsection{Achievable DoF over the Number of Antennas}
\begin{figure}[h]
  \centering
  \includegraphics[width=\columnwidth]{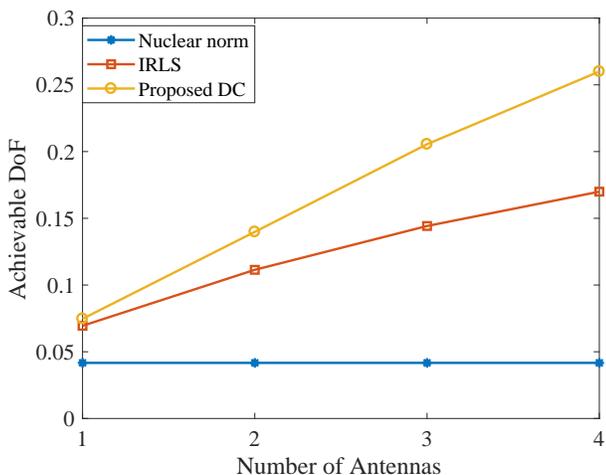}
  \caption{The maximum achievable symmetric DoF over the number of antennas when the mobile users and the AP are equipped with same number of antennas.}
  \label{fig:antenna}
\end{figure}

We consider a wireless distributed computing framework with $8$ mobile users and an AP. Each mobile user stores $1$ out of $4$ files in its local memory. We assume that each mobile users and the AP are equipped with the same number of antennas. We used different number of antennas to evaluate the multiplex gain of the focused wireless distributed computing system. Each point is averaged $100$ times and the result is shown in Fig. \ref{fig:antenna}. 

We can see that achievable symmetric DoF grows linearly with the number of antennas for the proposed DC algorithm and IRLS algorithm. However, the achievable DoF by the nuclear norm relaxation algorithm remains constant despite the growing number of antennas due to the poor structure of our problem. This test demonstrates that the proposed transceiver design framework achieves linear gain by increasing the number of antennas for the proposed DC algorithm. It also shows the intrinsic defects of nuclear norm relaxation approach for the data shuffling problem. The proposed DC approach is superior to the IRLS algorithm and the nuclear relaxation approach for data shuffling.

\subsection{Achievable DoF over the Number of Mobile Users}
\begin{figure}[h]
  \centering
  \includegraphics[width=\columnwidth]{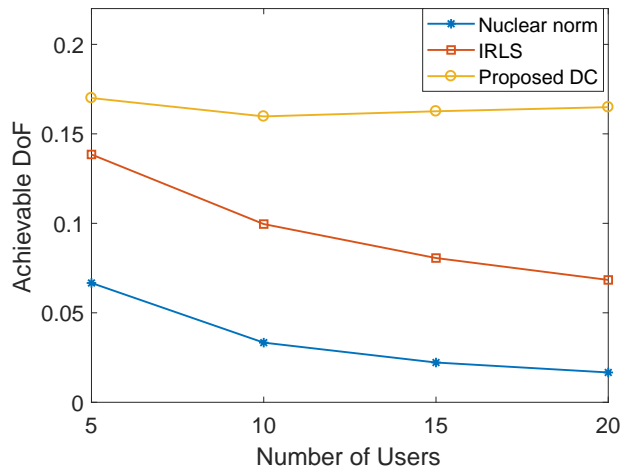}
  \caption{The achievable DoF with different algorithms over the number of mobile users.}
  \label{fig:users}
\end{figure}

As pointed in \cite{Li2017ascalable}, the limited communication bandwidth may become the bottleneck since the computation tasks increase linearly with network size. Therefore, the scalability becomes critical for a wireless distributed computing framework. In this test, we shall evaluate the achievable DoF by increasing the number of mobile users. Consider a single-antenna wireless distributed computing system where the dataset can be separated to $5$ files, and each mobile user can only store up to $2$ files in its local storage. We consider the uniform placement case when each mobile user stores $\mu=2$ files and each file is stored by $\mu K/N=2K/5$ mobile users. Consider the single datastream case of $d=1$. The achievable symmetric DoFs of different algorithms averaged over $100$ trials are shown in Fig. \ref{fig:users}. The achievable DoFs of the proposed DC algorithms remain nearly unchanged as the network size grows, which demonstrates its scalability. On the contrary, there is a marked decline of the achievable DoFs for IRLS algorithm and nuclear norm relaxation algorithm. Although more requested messages are involved in the system when the number of users grows, opportunities of collaboration for mobile users also increase since each file is stored at more mobile users. Our proposed algorithm can harness the benefits of such collaboration while other algorithms fail. However, it still remains an interesting but challenging problem to prove the scalability theoretically for the proposed DC algorithm.

In summary, the proposed DC algorithm has the capability of achieving higher DoF over benchmark approaches by exploiting the special structure of the data shuffling problem. Furthermore, the achievable DoF of the proposed DC algorithm almost remains unchanged when the number of mobile users increases.

\section{Conclusion}
In this paper we proposed a novel low-rank optimization to improve the communication efficiency for wireless distributed computing. We focus on the data-shuffle phase of the distributed computing and establish a novel interference alignment condition for data shuffling. We proposed a novel DC representation for the rank function based on Ky Fan $2$-$k$ norm, and then developed an efficient DC algorithm for the focused low-rank optimization problem, by deriving the closed-form solution for each iteration of the proposed DC algorithm. Numerical results demonstrated that the proposed DC approach can achieve higher DoF than the nuclear norm relaxation approach and IRLS algorithm. Furthermore, in uniform placement scenario, the achievable DoF nearly remains unchanged though more mobile users are involved. 

For the proposed data shuffling strategy for wireless distributed computing, there still exist some open problems. Possible future directions are listed as follows:
\begin{itemize}
  \item Although we have shown that the proposed low-rank approach is scalable with the growth of mobile users, it is particularly interesting to prove the scalability theoretically.
  \item We have shown that the proposed DC algorithm converges globally, but establishing the convergence rate can be considered in future works.
  \item It would also be interesting to consider the transceiver design with finite SNR scenarios for the proposed communication model for data shuffling in wireless distributed computing systems.
\end{itemize}

\appendices
\section{Proofs of Proposition \ref{prop:convergence}: Convergence of Algorithm \ref{algorithm:Proposed_DC}}\label{append:convergence}

Since $\bs{Y}^{[t]} \in \partial h(\bs{X}^{[t]})$, we have
\begin{equation}
	h(\bs{X}^{[t+1]})\geq h(\bs{X}^{[t]})+ \langle  \bs{X}^{[t+1]}-\bs{X}^{[t]}, \bs{Y}^{[t]} \rangle.
\end{equation}
Hence, it follows
\begin{equation}
	(g-h)(\bs{X}^{[t+1]}) \leq g(\bs{X}^{[t+1]})- \langle \bs{X}^{[t+1]} -\bs{X}^{[t]},\bs{Y}^{[t]} \rangle-  h(\bs{X}^{[t]}). \label{append:conv_eq1}
\end{equation}
Similarly, $\bs{X}^{[t+1]}\in \partial g^*(\bs{Y}^{[t]})$ implies
\begin{equation}
	g(\bs{X}^{[t]})\geq g(\bs{X}^{[t+1]})+ \langle  \bs{X}^{[t]}-\bs{X}^{[t+1]}, \bs{Y}^{[t]} \rangle+\|\bs{X}^{[t+1]}-\bs{X}^{[t]}\|_F^2.
\end{equation}
Thus, we obtain inequality
\begin{align}
	g(\bs{X}^{[t+1]})&-\langle \bs{X}^{[t+1]}-\bs{X}^{[t]}, \bs{Y}^{[t]} \rangle-h(\bs{X}^{[t]}) \nonumber\\& \leq (g-h)(\bs{X}^{[t]})-\|\bs{X}^{[t+1]}-\bs{X}^{[t]}\|_F^2. \label{append:conv_eq2}
\end{align}
On the other hand, 
\begin{align}
  \bs{X}^{[t+1]}\in \partial g^*(\bs{Y}^{[t]}) &\Leftrightarrow \langle\bs{X}^{[t+1]},\bs{Y}^{[t]}\rangle = g(\bs{X}^{[t+1]})+g^*(\bs{Y}^{[t]}) \\
  \bs{Y}^{[t]}\in \partial h(\bs{X}^{[t]}) &\Leftrightarrow \langle\bs{X}^{[t]},\bs{Y}^{[t]}\rangle = h(\bs{X}^{[t]})+h^*(\bs{Y}^{[t]}). \label{appdeq:eq1}
\end{align}
Then it follows
\begin{align}
  g(\bs{X}^{[t+1]})-\langle \bs{X}^{[t+1]} -\bs{X}^{[t]},&\bs{Y}^{[t]} \rangle-  h(\bs{X}^{[t]})  \nonumber\\
   &= h^*(\bs{Y}^{[t]})-g^*(\bs{Y}^{[t]}).
\end{align}

According to (\ref{append:conv_eq1}) and (\ref{append:conv_eq2}), we obtain that
\begin{align}
  \!\!(g-h)(\bs{X}^{[t+1]}) &\leq h^*(\bs{Y}^{[t]})-g^*(\bs{Y}^{[t]}) \nonumber \\
  &\leq (g-h)(\bs{X}^{[t]})-\|\bs{X}^{[t+1]}-\bs{X}^{[t]}\|_F^2.
\end{align}
Adding that
\begin{equation}
	(g-h)(\bs{X})\geq 0,
\end{equation}
the objective value converges and
\begin{equation}
  \lim_{t\rightarrow \infty}\|\bs{X}^{[t+1]}-\bs{X}^{[t]}\|_F^2=0.
\end{equation}
For every limit point,
\begin{equation}
  (g-h)(\bs{X}^{[t+1]}) = (g-h)(\bs{X}^{[t]}),
\end{equation}
and
\begin{equation}
  \|\bs{X}^{[t+1]}-\bs{X}^{[t]}\|_F^2=0.
\end{equation}
Therefore, we have 
\begin{align}
  (g-h)(\bs{X}^{[t+1]}) = h^*(\bs{Y}^{[t]})-g^*(\bs{Y}^{[t]}) = (g-h)(\bs{X}^{[t]}).
\end{align}
From (\ref{appdeq:eq1}) we know that
\begin{equation}
  h(\bs{X}^{[t+1]})+h^*(\bs{Y}^{[t]})=g(\bs{X}^{[t+1]})+g^*(\bs{Y}^{[t]}) =  \langle\bs{X}^{[t+1]},\bs{Y}^{[t]}\rangle,
\end{equation}
i.e.,
\begin{equation}
  \bs{Y}^{[t]}\in \partial h(\bs{X}^{[t+1]}).
\end{equation}
Then we have $\bs{Y}^{[t]}\in \partial g(\bs{X}^{[t+1]})\cap \partial h(\bs{X}^{[t+1]})$, which implies that $\bs{X}^{[t+1]}$ is a critical point of $g-h$. Therefore, given $r$ Algorithm \ref{algorithm:Proposed_DC} converges to critical points from arbitrary initial points.
\bibliographystyle{IEEEtran}
\bibliography{dist_comp}

\end{document}